\documentclass[11pt,reqno]{amsart}

\usepackage{amssymb}

\newcommand{\metric}{\mathrm{g}}

\begin{document}

\title[]{Scale Invariant Spectrum from Variable Speed of Light Metric in a Bimetric Gravity Theory}%

\author{M.~A.~Clayton}%
\address[1]{Department of Physics \\
Acadia University \\
Wolfville, Nova Scotia \\
B4P 2R6, Canada}
\email{michael.clayton@acadiau.ca}
\author{J.~W.~Moffat}%
\address[2]{Department of Physics \\
University of Toronto \\
Toronto, Ontario \\
M5S 1A7, Canada\\
and\\
The Perimeter Institute for Theoretical Physics\\
Waterloo, Ontario, N2J 2W9, Canada} \email{john.moffat@utoronto.ca}

\date{\today}

\thanks{PACS: 98.80.C; 04.20.G; 04.40.-b}

\keywords{Cosmology, causality, inflation, alternative theories of
gravity}


\begin{abstract}
An approximately scale invariant spectrum generating the seeds of structure formation is derived from a bimetric gravity theory.
By requiring that the amplitude of the CMB fluctuations from the model matches the observed value, we determine the fundamental length scale in the model to be a factor of $10^5$ times larger than the Planck length, which results in a scalar mode spectral index: $n_s\approx 0.97$, and its running:  $\alpha_s\approx -5\times 10^{-4}$. 
This is accomplished in the variable speed of light (VSL) metric frame, in which the dynamics of perturbations of the bimetric scalar field are determined by a minimally-coupled Klein-Gordon equation, and it is assumed that modes are born in a ground state at a scale given by the fundamental length scale appearing in the bimetric structure.
We show that while this is taking place for scales of interest, the background (primordial) radiation energy density is strongly suppressed as a result of the bimetric structure of the model.
Nevertheless, the enlarged lightcone of matter fields ensures that the horizon and flatness problems are solved.
\end{abstract}

\maketitle

\section{Introduction}

In previous published
work~\cite{Clayton+Moffat:1999a,Clayton+Moffat:2001,Clayton+Moffat:2002a,Moffat:2002},
a bimetric gravity theory was developed based on the geometric relationship:
\begin{equation}\label{eq:bimetric}
\hat{\metric}_{\mu\nu}=\metric_{\mu\nu}+B\partial_\mu\phi\partial_\nu\phi,
\end{equation}
with $B$ a constant with the dimensions of a squared length, and the scalar field
$\phi$ we shall call the biscalar field. The essential idea is that the tensor
$\hat{\metric}_{\mu\nu}$--which we refer to as the ``matter metric''--provides
the geometry on which matter fields propagate and interact, whereas it is the tensor
$\metric_{\mu\nu}$--which we refer to as the gravitational metric--that
represents the gravitational geometry through which gravitational waves propagate.
As we shall show in Section~\ref{sect:model intro}, the presence of this type of
prior-geometric structure in spacetime implies a differing propagation speed for matter and gravitational waves, and dramatically alters the coupling of matter to geometry. 
(An alternative bimetric theory has also been developed based on the structure: ${\hat g}_{\mu\nu}=g_{\mu\nu}+B\psi_\mu\psi_\nu$ where $\psi_\mu$ is a vector field~\cite{Clayton+Moffat:1999,Clayton+Moffat:2000}.)
While such structures have been considered by other authors~\cite{Beckenstein:1993,Beckenstein+Sanders:1994,Will:1993}, to our knowledge their consequences had not been considered in a cosmological setting prior to our work~\cite{Clayton+Moffat:1999a,Drummond:2000,Bassett+:2000,Clayton+Moffat:2001,Clayton+Moffat:2002a,Moffat:2002}.
It is also the explicit bimetric structure~\eqref{eq:bimetric} that distinguishes this approach from other investigations of models with a `varying speed of light' (see, for example~\cite{Moffat:1993a,Albrecht+Maguiejo:1999}).

A derivation of an approximately scale-invariant microwave background spectrum was
obtained previously~\cite{Clayton+Moffat:2002a} by invoking a secondary mechanism,
similar to that of chaotic inflation, whereby a mass was introduced for the biscalar
field and the quantum fluctuations in the biscalar field were generated as the
biscalar field `rolled' along the attractor as it approached the bottom of the
potential well. However, it is desirable to discover whether an alternative `purely'
VSL mechanism exists which can generate a scale invariant
microwave background spectrum, without relying on the introduction of a potential,
and therefore bears no relation to a potential-dominated inflationary scenario.

In the following, we will demonstrate in the affirmative that this is indeed the
case. In Section~\ref{sect:FRW}, we introduce a frame where the gravitational metric
$\metric_{\mu\nu}$ takes on the standard comoving Friedman-Robertson-Walker form,
which we will refer to as the ``VSL frame'', since in this frame the speed of
light will vary as the universe expands. In Section~\ref{sect:fluctuations}, we show
that perturbations of the scalar field $\phi$ that provide the bimetric structure
in~\eqref{eq:bimetric} satisfy an Einstein Klein-Gordon equation. With vanishing
potential for this field, we can obtain the desired scale invariant spectrum by
assuming that the scalar perturbation modes are born, or emerge from a spacetime
region with a length scale less than a fixed length scale $\ell_0$. 
This follows the idea of Hollands and
Wald~\cite{Hollands+Wald:2002a,Hollands+Wald:2002b} in which the modes emerge in a
ground state wavefunction at a scale $R_k=R(t_k)$ at a time $t_k$.

In contrast to the picture described by Hollands and Wald, the radiation density in
the volume of space determined by $\ell_0$ is highly diluted by means of a VSL
mechanism.
This justifies the use of the biscalar field to generate the seeds of structure
formation, circumventing the major criticism by Kofman, Linde and
Mukhanov~\cite{Kofman+Linde+Mukhanov:2002} of the Hollands and Wald scenario. 
The other important observation that we make here is that $\ell_0$ can be chosen to be related to the parameter $B$ in the bimetric structure through: $\ell_0\approx \sqrt{12B}$, and we obtain the correct order of magnitude of the CMB spectrum $\delta_H\approx 10^{-5}$ if $\ell_0\approx 10^5\ell_P$, where $\ell_P$ is the Planck length $\ell_P=\sqrt{G\hbar/c^3}\approx 10^{-33}\,\mathrm{cm}$. 
We therefore feel that the bimetric structure should also play a fundamental role in understanding the mode creation mechanism as well. Following a brief discussion of the consistency of choosing the parameters of the model in Section~\ref{sect:initial}, we make some concluding statements in Section~\ref{sect:conclusion}.

\section{The model}
\label{sect:model intro}

The model that we introduced in~\cite{Clayton+Moffat:1999a} consisted of a self-gravitating scalar
field coupled to matter through the matter metric~\eqref{eq:bimetric}, with the
action
\begin{equation}
S=S_{\rm grav}+S_{\phi}+\hat{S}_{\rm M},
\end{equation}
where
\begin{equation}
S_{\rm grav}=-\frac{1}{\kappa}\int d\mu (R[\metric]+2\Lambda),
\end{equation}
$\kappa=16\pi G/c^4$, $\Lambda$ is the cosmological constant, and
we employ a metric with signature $(+,-,-,-)$.  We will write, for
example, $d\mu=\sqrt{-\metric}\;d^4x$ and $\mu=\sqrt{-\metric}$
for the metric density related to the gravitational metric
$\metric_{\mu\nu}$, and similar definitions of $d\hat{\mu}$ and
$\hat{\mu}$ in terms of the matter metric $\hat{\metric}_{\mu\nu}$, and $c$ denotes
the currently measured speed of light. Note that from~\eqref{eq:bimetric} the
determinants are related by
\begin{equation}\label{eq:density relation}
\mu=\sqrt{K}\hat{\mu},
\end{equation}
where
\begin{equation}\label{eq:K defn}
K=1-B\hat{\metric}^{\mu\nu}\partial_\mu\phi\partial_\nu\phi.
\end{equation}

The minimally-coupled scalar field action is
\begin{equation} S_{\rm \phi}=\frac{1}{\kappa}\int d\mu\,
\Bigl[\frac{1}{2}\metric^{\mu\nu}\partial_\mu\phi\partial_\nu\phi-V(\phi)\Bigr],
\end{equation}
where the scalar field $\phi$ has been chosen to be dimensionless.
The energy-momentum tensor for the scalar field that we will use is given by
\begin{equation}
T^{\mu\nu}_\phi=\frac{1}{\kappa}\Bigl[
\metric^{\mu\alpha}\metric^{\nu\beta}\partial_\alpha\phi\partial_\beta\phi
-\tfrac{1}{2}\metric^{\mu\nu}\metric^{\alpha\beta}\partial_\alpha\phi\partial_\beta\phi
+\metric^{\mu\nu}V(\phi) \Bigr],
\end{equation}
and is the variation of the scalar field action with respect to the gravitational
metric: $\delta S_{\phi}/\delta \metric_{\mu\nu} =-\frac{1}{2}\mu T_\phi^{\mu\nu}$.

We construct the matter action $\hat{S}_{\mathrm{M}}$ using the
combination~\eqref{eq:bimetric} resulting in the identification of
$\hat{\metric}_{\mu\nu}$ as the physical metric that provides the arena on which
matter fields interact. That is, the matter action $\hat{S}_{\mathrm{M}}[\psi^I] =
\hat{S}_{\mathrm{M}}[\hat{\metric},\psi^I]$, where $\psi^I$ represents all the
matter fields in spacetime, is one of the standard forms, and therefore the
energy-momentum tensor derived from it by
\begin{equation}\label{eq:matterEM}
\frac{\delta S_{\mathrm{M}}}{\delta \hat{\metric}_{\mu\nu}}
 =-\frac{1}{2}\hat{\mu}\hat{T}^{\mu\nu},
\end{equation}
satisfies the conservation laws
\begin{equation}\label{eq:matterconservation}
\hat{\nabla}_\nu\Bigl[\hat{\mu}\hat{T}^{\mu\nu}\Bigr]=0,
\end{equation}
as a consequence of the matter field equations only~\cite{Clayton+Moffat:1999a}.
It is the matter covariant derivative $\hat{\nabla}_\mu$ that appears here, which is
the metric compatible covariant derivative determined by the matter metric:
$\hat{\nabla}_\alpha\hat{\metric}_{\mu\nu}=0$. In this work we assume a perfect
fluid matter model:
\begin{equation}
 \hat{T}^{\alpha\beta}=
 \Bigl(\rho+\frac{p}{c^2}\Bigr)\hat{u}^\alpha\hat{u}^\beta
 -p\hat{\metric}^{\alpha\beta},
\end{equation}
with $\hat{\metric}_{\mu\nu}\hat{u}^\mu\hat{u}^\nu=c^2$.

As described in~\cite{Clayton+Moffat:1999a}, the gravitational field equations for
this model can be written as
\begin{equation}\label{eq:Einsteins eqns}
 G^{\mu\nu}=\Lambda\metric^{\mu\nu}
 +\frac{\kappa}{2}T^{\mu\nu}_\phi
 +\frac{\kappa}{2}\frac{\hat{\mu}}{\mu}\hat{T}^{\mu\nu},
\end{equation}
and that for the scalar field (written here in terms of matter
covariant derivatives) as
\begin{equation}\label{eq:scalar FEQ}
\bar{\metric}^{\mu\nu}\hat{\nabla}_\mu\hat{\nabla}_\nu\phi+KV^\prime
[\phi]=0.
\end{equation}
In the latter, we have defined the biscalar field metric
\begin{equation}\label{eq:scalar metric}
 \bar{\metric}^{\mu\nu} = \hat{\metric}^{\mu\nu}
 +\frac{B}{K}\hat{\nabla}^\mu\phi\hat{\nabla}^\nu\phi
 -\kappa B\sqrt{K}\hat{T}^{\mu\nu},
\end{equation}
which is a third metric structure forced upon us by the structure of the field
equations, and controls the causal propagation of the biscalar field. We have
previously shown that the field equations in this theory are consistent with the
Bianchi identities~\cite{Clayton+Moffat:1999a}.

Note that there is no source for the biscalar field in~\eqref{eq:scalar FEQ}.
This implies that as the universe expands the biscalar field becomes increasingly diluted to the point where it has unobservable consequences at present, despite the rather large energy scale implied by the relation $\sqrt{12B}=\ell_0\sim 10^5\,\ell_P$ that we derive below.
To see this, perform a post-Newtonian expansion of the gravitational field in the solar system: $\mathrm{g}_{00}\sim c^2 + 2GM_\odot/1\,\mathrm{AU}$; if the contribution from the biscalar field is to be as large as the Newtonian gravitational potential, then we have: $2GM_\odot/1\,\mathrm{AU}\sim B\dot\phi^2$.
Using the determined magnitude of $B$ this can be written: $\dot{\phi}^2=24 M_\odot c^3 /(10^{10}\hbar\,1\,\mathrm{AU})$ or $\dot{\phi}\sim 10^{35}\,\mathrm{Hz}\sim 10^{-8}/t_P$.
In the very early universe the biscalar field will be `rolling' fast enough to approach these energies, but as it will redshift as $\dot\phi\sim 1/R^3$ (see~\eqref{eq:dot phi solution}), such biscalar field effects will be unobservable at present.

\section{Cosmology}
\label{sect:FRW}

We will work in a frame where the gravitational metric $\metric_{\mu\nu}$ is comoving, that is
\begin{equation}\label{eq:gravmetric}
{\metric}_{\mu\nu}=\mathrm{diag}(c^2,-R^2(t)\gamma_{ij}),
\end{equation}
with coordinates $(t, x^i)$ and $3$-metric $\gamma_{ij}$ on the
spatial slices of constant time.
As a result of this coordinate choice and the definitions~\eqref{eq:bimetric}
and~\eqref{eq:scalar metric}, we find
\begin{equation}\label{eq:FRW ghat}
 \hat{\metric}_{\mu\nu}=\mathrm{diag}(K^{-1}c^2,-R^2(t)\gamma_{ij}),
\end{equation}
where from~\eqref{eq:K defn}:
\begin{equation}
 K = \left( 1 + \frac{B}{c^2}\dot{\phi}^2\right)^{-1}.
\end{equation}
From~\eqref{eq:gravmetric} we see that the constant $c$ will represent the speed of
propagation of gravitational waves in this model, whereas from~\eqref{eq:FRW ghat}
the speed of light (and the limiting speed of other matter fields) is time-dependent
and given by
\begin{equation}
c_\gamma(t)=c/\sqrt{K}=c\left(1+\frac{B}{c^2}\dot{\phi}^2\right)^{1/2}.
\end{equation}
We see that $c_\gamma(t)\rightarrow\infty$ and $K(t)\rightarrow 0$ when $B\dot\phi^2/c^2\rightarrow\infty$, and expect that at present $K\sim 1$ and so $c_\gamma\sim c$.

In general relativity (GR), one can always perform a diffeomorphism to remove such a
time-dependence of the speed of light. In the bimetric gravity theory there are two
speeds, and, while this can still be achieved by introducing a new time variable
through $d\tau = dt/\sqrt{K}$, in the resulting frame the speed of gravitational
wave propagation would no longer be constant. Note that the dimensionless ratio of
the two propagation velocities: $c/c_\gamma = \sqrt{K}$ is a covariantly defined
quantity by~\eqref{eq:K defn}, and encodes exactly this situation: only if $K=1$
will a diffeomorphism be able to simultaneously make both propagation velocities
equal. This makes the time dependence of $c_\gamma$ (in this frame) a non-trivial
feature of the theory.

The matter stress-energy tensor (using $\hat{u}^0=c$):
\begin{equation}
 \hat{T}^{00}=K\rho,\quad
 \hat{T}^{ij}=\frac{p}{R^2}\gamma^{ij},
\end{equation}
then leads to the conservation laws (an overdot indicates a
derivative with respect to the time variable $t$, and
$H=\dot{R}/R$ is the Hubble function)
\begin{equation}\label{eq:cosm cons}
\dot{\rho}+3H\Bigl(\rho+\frac{p}{c^2}\Bigr)=0.
\end{equation}
Since we are interested in the very early universe, we will assume
a radiation equation of state:
\begin{equation}\label{eq:equation of state}
p=\frac{1}{3}c^2\rho,
\end{equation}
and we will write:
\begin{equation}
\rho = \rho_{\mathit{rad},0}\left(\frac{R_0}{R}\right)^4,
\end{equation}
where the radiation energy density at present is: $\rho_{\mathit{rad},0}\approx
4.8\times 10^{-34}\,\mathrm{g}/\mathrm{cm}^3$. In this model we are assuming that
the radiation that fills the universe is truly `primordial'.

The Friedmann equation is given by
\begin{equation}\label{eq:Fried eqn}
 H^2 +\frac{kc^2}{R^2} =
 \frac{1}{3}c^2\Lambda
+\frac{1}{6}\left(\frac{1}{2}\dot{\phi}^2+c^2V[\phi]\right)
 +\frac{1}{6}\kappa c^4 \sqrt{K}\rho,
\end{equation}
and the biscalar field equation is:
\begin{equation}\label{eq:biscalar eqn}
 (1-\kappa c^2 B K^{3/2}\rho)\ddot{\phi}
 +3H( 1+ \kappa B \sqrt{K}p)\dot{\phi}
 +c^2 V^\prime[\phi]
 =0,
\end{equation}
with the biscalar field metric given by:
\begin{equation}
 \bar{\metric}_{\mu\nu}
=\mathrm{diag}(c^2(1-\kappa c^2 B K^{3/2}\rho),
-R^2(t)(1+\kappa B \sqrt{K} p)\gamma_{ij}).
\end{equation}
It is useful at this point to introduce the following quantities
derived from the constant $B$ which will appear throughout this
work:
\begin{equation}\label{eq:definitions}
H_B^2=\frac{c^2}{12B},\quad
\rho_B=\frac{1}{2\kappa c^2 B},
\end{equation}
the latter comes from $H_B^2=\frac{1}{6}\kappa c^4\rho_B$.

Previously~\cite{Clayton+Moffat:2002a}, we considered the generation of the seeds for
structure formation in the early universe through a mechanism similar
to that of chaotic inflation. We shall now obtain the seeds for structure
formation from a purely VSL mechanism by working in the comoving gravitational
metric (\ref{eq:gravmetric}) (VSL metric frame). The essential idea here is that it is
possible in the very early universe that $K$ is very small: $B\dot{\phi}^2\gg c^2$,
in which case the matter contribution to the Friedmann equation is small compared to the contribution from the biscalar field kinetic term.
We are interested in a scenario that does \textit{not} depend on
choosing a particular form for the biscalar field potential, and so we will set
$V[\phi]=0$.

We assume in~\eqref{eq:biscalar eqn} that the bracketed terms are both approximately
one, and so: $\bar{\metric}_{\mu\nu}\approx \metric_{\mu\nu}$ (we shall justify this assumption in Section~\ref{sect:initial}), which gives us the solution:
\begin{equation}\label{eq:dot phi solution}
\dot{\phi}=\sqrt{12}H_B\left(\frac{R_{\mathit{pt}}}{R}\right)^3.
\end{equation}
We have chosen the arbitrary constant of integration $R_{\mathit{pt}}$ to parameterize the
time at which $K=1/2$, that is
\begin{equation}\label{eq:K soln}
K=\left[1+\left(\frac{R_{\mathit{pt}}}{R}\right)^6\right]^{-1},
\end{equation}
so that it indicates the time at which the gravitational and matter metrics are close to coinciding.
The subscript $\mathit{pt}$ indicates that this is the `phase transition', when
standard local Lorentz symmetry with a single light cone will  be restored--it is
also the end of the period that will appear as inflation in the matter frame.

The Friedmann equation is then
\begin{equation}\label{eq:reduced Fried eqn}
 H^2 +\frac{kc^2}{R^2} =
 \frac{1}{3}c^2\Lambda
+H_B^2\left(\frac{R_{\mathit{pt}}}{R}\right)^6,
\end{equation}
from which it is easy to see that at early enough times, the behavior of the universe is dominated by the contribution from the biscalar field.
We shall assume that this is the case, and consequently neglect the cosmological constant and spatial curvature throughout the remainder of this work.
As a result, we have the approximate solution
\begin{equation}\label{eq:Rsoln}
R= R_{\mathit{pt}}(3H_Bt)^{1/3},\quad
H=\frac{1}{3t},
\end{equation}
where we have chosen $t=0$ when $R=0$, and at the phase transition $t_{\mathit{pt}}=1/(3H_B)$, which means that $H_{\mathit{pt}}=H_B$.

As this is not an inflationary solution, it appears that it does not solve the horizon and flatness problems.
It must be remembered though, that we are working in a comoving gravitational frame, and the speed of matter propagation is very much larger than that of gravitational waves.
So the coordinate distance that light would travel since the initial singularity is given by the null geodesics of~\eqref{eq:FRW ghat} rather than those of~\eqref{eq:gravmetric}, that is: 
\begin{equation}
\int \frac{c\,dt}{KR}
\propto \frac{1}{t^{4/3}},
\end{equation}
where we have used~\eqref{eq:Rsoln}.
This clearly diverges as $t\rightarrow 0$, showing that there is no (matter) particle horizon in the spacetime.

Another way of seeing this is to perform the diffeomorphism to put the matter metric into comoving form:
\begin{equation}\label{eq:comoving matter time}
d\tau
= K^{-1/2}dt
= \sqrt{1+(3H_Bt)^{-2}}dt,
\end{equation}
which is easily integrated:
\begin{equation}
3H_B\tau = -1 + \sqrt{1+(3H_Bt)^2}
+\frac{1}{2}\ln\left[\frac{\sqrt{1+(3H_Bt)^2}-1}
{\sqrt{1+(3H_Bt)^2}+1}\right],
\end{equation}
where we have chosen the constant of integration so that $\tau=0$ corresponds to
$3H_B t\approx 1$.
Note that the interval $t\in (0,\infty)$ has been mapped to $\tau\in(-\infty,\infty)$, so a comoving matter observer will `see' an infinitely old universe.
Using this, it is not difficult to show that $t\approx \tau$ for $3H_B t \gg 1$ and $3H_B t \approx 2\exp(3H_B \tau)$ for $3H_B t \ll 1$.
Thus, in the comoving matter frame we have
\begin{equation}
R(\tau)\approx \begin{cases}
R_{\mathit{pt}} 2^{1/3}e^{3H_B\tau} & 3H_B t \lesssim 1,\\
R_{\mathit{pt}}(3H_Bt)^{1/3} & 3H_B t \gtrsim 1,
\end{cases}
\end{equation}
which clearly shows that a comoving material observer will see a period of inflation.
These results can be understood directly from the Friedmann equations in a comoving matter frame as well~\cite{Clayton+Moffat:2001}.

This is a recurring theme in such bimetric models: the frame in which one chooses to work colours the interpretation that one puts on the physics.
In the comoving gravitational frame the speed of light varies in time, and the horizon problem is solved due to the very large propagation velocity of light in the very early universe.
In a frame where the speed of light is constant, the altered coupling of matter to the gravitational sector and the presence of the biscalar field leads to an inflating spacetime.

\section{Scalar Mode Perturbations}
\label{sect:fluctuations}

In the comoving gravitational frame we are considering a universe with a
minimally-coupled Einstein Klein-Gordon field, and so scalar mode fluctuations: $\phi=\phi(t)+\delta\phi(t,\vec{x})$, about the background cosmological solution can be treated in a straightforward way.
Ignoring gravitational back-reaction, we find the field equation for the
perturbation of the biscalar field
\begin{equation}\label{eq:phi pert}
\frac{d^2\delta\phi_{\vec{k}}}{dt^2}
+3H\frac{d \delta\phi_{\vec{k}}}{dt}
+\frac{c^2\vec{k}^2}{R^2}\delta\phi_{\vec{k}}=0,
\end{equation}
where $\delta\phi  = (2\pi)^{-2/3}\int d^3k\,\exp[-i\vec{k}\cdot\vec{x}]\delta\phi_{\vec{k}}$.
The (real) solution to~\eqref{eq:phi pert} for our spacetime is given by Bessel functions:
\begin{equation}
\delta\phi_{\vec{k}} = A(\vec{k})J_0(\xi)+B(\vec{k})Y_0(\xi),\quad
\xi=\frac{ck}{2R_{\mathit{pt}}H_B}\left(\frac{R}{R_{\mathit{pt}}}\right)^2.
\end{equation}
Since $\xi\gtrsim 1$ would correspond to a mode passing inside the horizon, we expect that in the very early universe we would have $\xi\ll 1$ for modes of interest.
Using the fact that $R_0H_0/(R_{\mathit{pt}}H_{\mathit{pt}})\ll 1 $, for modes near the pivot point: $ck\sim 7 R_0H_0$ we see that this is the case.
Thus, we cannot assume a scenario where the modes are `born' in the quantum vacuum in the early universe.

Also note that re-writing~\eqref{eq:phi pert} in terms of the matter comoving time $\tau$ as defined by~\eqref{eq:comoving matter time} results in a field equation that is very different than in what would appear in a deSitter spacetime:
\begin{equation}
\frac{d^2\delta\phi_{\vec{k}}}{d\tau^2}
+\frac{d\ln\left(R^3/\sqrt{K}\right)}{d\tau}\frac{d \delta\phi_{\vec{k}}}{d\tau}
+\frac{Kc^2\vec{k}^2}{R^2}\delta\phi_{\vec{k}}=0.
\end{equation}
The additional factor $K$ multiplying the spatial derivative terms is a reflection of the fact that the null cone of the biscalar field is smaller by exactly this factor over that of matter.
It is also easy to see that since in the very early universe $K\sim R^6$, the biscalar field modes never appear to be `inside the horizon'.

Hollands and Wald~\cite{Hollands+Wald:2002a} assume that modes are born or emerge from a fundamental description of spacetime in the `ground state' of a flat spacetime wave operator at some length scale $\ell_0$.
We interpret this to mean that modes emerge from a spacetime foam in a classical state that is essentially a normalized plane wave of the flat spacetime wave operator:
\begin{equation}
\label{eq:classicalwave}
\delta\phi_{\vec{k}}(t_k) = \sqrt{\frac{\kappa \hbar c^2}{(2\pi R_k)^32\omega_k}}
\cos(\omega_k t_k -\vec{k}\cdot\vec{x} +\delta),\quad
\omega_k =\frac{ck}{R_k},
\end{equation}
where the scale at which the mode is born is given by the Hollands-Wald condition:
\begin{equation}\label{eq:HW condition}
R_k=k\ell_0.
\end{equation}
Since we will eventually relate this length scale to that appearing in the metric~\eqref{eq:bimetric} by $\sqrt{12B}\sim \ell_0$, we expect that a more fundamental description of this mechanism in terms of (quantum) biscalar field dynamics should be possible.
We leave this to future work.

If one assumes that one should use~\eqref{eq:classicalwave} as initial data for the classical solution of~\eqref{eq:phi pert}, then we match not only the initial perturbation but also its time derivative. 
Doing so and keeping only the dominant contribution as $\xi\rightarrow 0$ gives
\begin{equation}
\delta\phi_{\vec{k}} \approx
\sqrt{\frac{9\kappa \hbar c^2}{(2\pi R_k)^3 32\omega_k}}
\cos(\omega_k t_k -\vec{k}\cdot\vec{x} +\delta)
\ln(\xi_k)
J_0(\xi),
\end{equation}
where $\xi_k$ represents $\xi$ evaluated at $R=R_k$.
The (non-scale invariant) contribution that appears here results from the fact that we matched to the initial state and its time derivative, and the Bessel function $Y_0(z)$ is logarithmically divergent when $\xi\rightarrow 0$. 
The Hollands and Wald~\cite{Hollands+Wald:2002a} wave function did not have this additional logarithmic term, which will lead to slight deviations from a scale invariant spectrum.

We obtain the spectrum of scalar field fluctuations to be
\begin{equation}
\mathcal{P}_{\delta\phi}
=\frac{9}{2(2\pi)^3}\left(\frac{\ell_P}{\ell_0}\right)^2
\ln^2(\xi_k),
\end{equation}
and the curvature power spectrum is then found in the usual way (recall that
$\dot{\phi}^2=12 H^2$ in this spacetime):
\begin{equation}\label{eq:PR}
\mathcal{P}_{\mathcal{R}}
=\biggl(\frac{H^2}{\dot{\phi}^2}\biggr)\mathcal{P}_{\delta\phi}
=\frac{3}{8(2\pi)^3}\left(\frac{\ell_P}{\ell_0}\right)^2
\ln^2(\xi_k),
\end{equation}
where, using the condition~\eqref{eq:HW condition}:
\begin{equation}\label{eq:xi k}
\xi_k=\frac{\sqrt{12B}\ell^2_0 k^3}{2R_{\mathit{pt}}^3}.
\end{equation}

From~\eqref{eq:PR} the scalar mode spectral index is calculated:
\begin{equation}
n_s=1+\frac{d\ln \mathcal{P}_{\mathcal{R}}}{d\ln k}
=1+\frac{6}{\ln(\xi_k)},
\end{equation}
and the running of the spectral index is calculated from: $\alpha_s=dn_s/d\ln k$, from which we find the relation
\begin{equation}
\alpha_s=-\tfrac{1}{2}(1-n_s)^2.
\end{equation}
In the large scale limit, we also have
\begin{equation}\label{eq:delta H}
\delta_H= \frac{2}{5}\sqrt{\mathcal{P}_{\mathcal{R}}}
\approx
\frac{2}{5}\sqrt{\frac{3}{8(2\pi)^3}}\biggl(\frac{\ell_P}{\ell_0}\biggr)
|\ln(\xi_k)|.
\end{equation}
Assuming that $\ell_0\approx \sqrt{12B}$ and evaluating the logarithm at the pivot point, $c k \sim 7 R_0H_0$, this simplifies to
\begin{equation}
\delta_H\approx \frac{\ell_P}{\ell_0} .
\end{equation}
To see this, note that for this scale: $\xi_k\approx (7 H_0R_0/H_{\mathit{pt}}R_{\mathit{pt}})^3$, where we have used: $H_{\mathit{pt}}=H_B$ and~\eqref{eq:definitions}.
Using standard physics following $R_{\mathit{pt}}$, one finds that $H_{\mathit{pt}}R_{\mathit{pt}}/R_0 H_0\approx \sqrt{10^{60} z_{\mathrm{eq}}}(T_P/T_{\mathrm{pt}})$~\cite{Coles+Lucchin:1995}, which gives $\ln(\xi_k)\approx -200$, which combines with the coefficient in~\eqref{eq:delta H} to give a factor of order unity.
We can then match the amplitude of the observed CMB fluctuations: $\delta_H\sim 10^{-5}$ by fixing the bimetric length scale: 
\begin{equation}\label{eq:l0 constraint}
\ell_0\approx \sqrt{12B} \approx 10^5 \, \ell_P,
\end{equation}
which gives
\begin{equation}
n_s \approx 0.97,\quad
\alpha_s \approx -4.5 \times 10^{-4}.
\end{equation}
This compares well with the recent WMAP result: $n_s=0.99\pm 0.04$~\cite{Bennett+etal:2003a}.

While it is straightforward to use this mechanism to also generate a nearly scale
invariant spectrum of tensor fluctuations, we have to show that the ratio of the
scalar and tensor amplitudes is relatively small to match the WMAP
data. The observational upper bound on this
ratio is of order 20\%~\cite{Bennett+etal:2003a,Bennett+etal:2003b}. This we leave
for future work.

\section{The Very Early Universe}
\label{sect:initial}

Now that we have determined a reasonable choice of parameters, we can justify the assumption that we made in Section~\ref{sect:FRW}, namely, that we can take $\bar{\metric}_{\mu\nu}\approx \metric_{\mu\nu}$.
In order to do so it is sufficient to show that $\kappa c^2 B \sqrt{K}\rho \ll 1$ since from~\eqref{eq:K soln}, $K\ll 1$ in the very early universe and we are assuming the equation of state~\eqref{eq:equation of state}.

Using the Friedmann equation~\eqref{eq:Fried eqn} and the biscalar field
equation~\eqref{eq:biscalar eqn}, it is straightforward to show (with $V=0$) that
\begin{equation}
\frac{d}{dt}\left[1-\kappa c^2B \sqrt{K}\rho\right]
=3H\kappa c^2 B K^{3/2}\left(\rho+K^{-1}\frac{p}{c^2}\right)
\frac{1-\kappa c^2B \sqrt{K}\rho}{1-\kappa c^2B K^{3/2}\rho},
\end{equation}
and so $\kappa c^2B \sqrt{K}\rho=1$ is a fixed point of the system.
It is also a degeneracy point of the biscalar field metric though, and is the $R\rightarrow 0$ limit of any universe in which $\rho$ is non-vanishing.
So if we were to follow the model described in Section~\ref{sect:FRW} back to $R=0$, there would necessarily be an earlier time period during which matter fields would be important.

Rather than dominating the Friedmann equation, the presence of matter merely serves
to cause $K\sim R^4$ rather than $K\sim R^3$ as $R\rightarrow
0$~\cite{Clayton+Moffat:2001}. This is easily seen from the Friedmann
equation~\eqref{eq:Fried eqn}, where the matter energy density contributions
approach a constant $\propto H_B^2$ whereas the biscalar field kinetic terms
continue to dominate: $\dot{\phi}^2\sim H_B^2/K$. We will not treat this stage of
the universe explicitly. Instead we assume that modes of interest are generated
after this stage, and so if we define the quantity:
\begin{equation}\label{eq:F defn}
F=2\kappa c^2B \sqrt{K}\rho
= \sqrt{K}\frac{\rho}{\rho_B},
\end{equation}
we should have $F\lesssim 1/2$.

The transition $F\approx 1$ should happen for some $R_i \ll R_{\mathit{pt}}$, and
using~\eqref{eq:K soln} this initial transition is given by
\begin{equation}\label{eq:Ri}
\frac{R_i}{R_{\mathit{pt}}}
= e^{-\mathcal{N}}
\approx \frac{\rho_{\mathit{rad},0}}{\rho_B}
\left(\frac{R_0}{R_{\mathit{pt}}}\right)^4
=\frac{\rho_P}{\rho_B}
\left(\frac{T_{\mathit{pt}}}{T_P}\right)^4 ,
\end{equation}
where we have introduced the total number of e-folds of inflation $\mathcal{N}$ in
the comoving matter frame, and we have used $\rho_PR_P^4=\rho_{\mathit{rad},0}R_0^4$, $R_0T_0=RT$ and $\rho_P=c^5/(G^2\hbar)$ is the Planck mass density.

In order to solve the horizon and flatness problems in the comoving matter frame, we
need the size of the cosmological horizon: $r_c=cR_0/(RH)$ to be much larger at
$R_{\mathit{qg}}$ than it is at present, leading to $R_0H_0\gg
R_{\mathit{qg}}H_{\mathit{qg}}$. Relating these to the phase transition (the end of
inflation in the comoving matter frame) we have
\begin{equation}
\frac{R_0H_0}{R_{\mathit{pt}}H_{\mathit{pt}}}\gg
\frac{R_iH_i}{R_{\mathit{pt}}H_{\mathit{pt}}}
\approx e^{-\mathcal{N}},
\end{equation}
where we have noted that the Hubble parameter is approximately constant during the
early universe in the comoving matter frame. Using standard physics following
$R_{\mathit{pt}}$ interpreted as the end of `inflation', this yields the constraint~\cite{Coles+Lucchin:1995}:
\begin{equation}
e^\mathcal{N} \gg \frac{T_{\mathit{pt}}}{T_P}\sqrt{10^{60}\,z_{eq}},
\end{equation}
where we have used $R_{\mathit{pt}}H_{\mathit{pt}}/(R_0H_0)=\sqrt{10^{60}z_{\mathrm{eq}}}(T_{\mathit{pt}}/T_P)$, and $T_{\mathit{pt}}$ is the temperature at the phase transition satisfying
$1>T_{\mathit{pt}}/T_P>10^{-5}$, $T_P$ is the Planck temperature, and
$z_{eq}\approx 4.3\times 10^4\,\Omega h\approx 4\times 10^4$ is the redshift at
matter-radiation equality. This yields the usual result that $\mathcal{N}\gtrsim 60$
will solve the horizon and flatness problems.

Evaluating~\eqref{eq:F defn} at the time when a particular mode was born, we have
\begin{equation}
F_k = \frac{4\pi}{3}\frac{12B}{\ell_P}
\left(\frac{T_{\mathit{pt}}}{T_P}\right)^4
\left(\frac{R_{\mathit{pt}}}{k\ell_0}\right),
\end{equation}
which for the pivot point, $ck\sim 7R_0H_0$, reduces to
\begin{equation}
F_{\mathrm{pivot}} \approx 2\times 10^{29}
\frac{(12B)^{3/2}}{\ell_0\ell_P^2}
\left(\frac{T_{\mathit{pt}}}{T_P}\right)^5,
\end{equation}
and using $l_0=\sqrt{12B}$ and~\eqref{eq:l0 constraint}, the condition
$F_{\mathrm{pivot}} \lesssim 1$ leads to
\begin{equation}\label{eq:Temperature inequality}
T_{\mathit{pt}}\lesssim 10^{-8}T_P \approx 10^{24}\,\mathrm{K}.
\end{equation}

Using this result in~\eqref{eq:Ri} we can determine the ratio:
\begin{equation}
\frac{\rho_B}{\rho_P}\lesssim 10^{-32} e^{\mathcal{N}},
\end{equation}
so that if we assume that the inequality~\eqref{eq:Temperature inequality} is
saturated, and provided that $\mathcal{N}\lesssim 74$ we will have $\rho_B < \rho_P$.

Thus, we have shown that modes of interest can be created during the time when
$F\lesssim 1$, that we have sufficient inflation in the matter metric frame to solve
the horizon and flatness problems, and that the fundamental length scale of our model
$\sqrt{B}$ is larger than the Planck length $\ell_P$.

\section{Conclusions}
\label{sect:conclusion}

We have sought to derive an approximately scale invariant CMB spectrum from a purely
VSL mechanism, working in a frame where the speed of light $c_\gamma(t)$ is
dynamical. In the limit of large $c_\gamma(t)$, we find that the radiation density
and entropy in the very early universe are significantly diluted, so that the
Einstein Klein-Gordon wave equation legitimately controls the behavior of the
perturbative scalar wave modes.

These wave modes are born in a spatial volume determined by the length scale
$\sqrt{12B}$ that appears in the bimetric structure of our theory, and are matched
to a state that is consistent with the flat space classical wave
function~\eqref{eq:classicalwave}. Thus, we do not obtain the seeds of structure
formation from quantum fluctuation modes in a de Sitter adiabatic vacuum, as in
the case of standard inflation models~\cite{Linde:1990}. The derivation of the scale
invariant spectrum from inflation models is not unique, in so far as one has to
choose the initial conditions for inflation either through the `new' inflationary
models or the `chaotic' models~\cite{Linde:1990}. The former depend on having an
initial universe that is sufficiently homogeneous to allow enough e-folds of
inflation, while the latter depend on an `anthropic' principle for their motivation.
Moreover, the choice of an adiabatic vacuum for the emergence of the fluctuation
modes in inflationary models is ambiguous~\cite{ArmendarizPicon+Lim:2003}. In our model, the
choice of initial conditions relies on $c_\gamma(t)$ becoming large in the early
universe, so that the VSL mechanism (or inflationary mechanism in the gravitational
metric frame) can become operative.

A calculation of the power spectrum leads to a fit to the thermal CMB
spectrum $\Delta T/T\sim 10^{-5}$ for the length scale $\sqrt{12B}\sim 10^5\ell_P$
and the prediction for the spectrum index $n_s\sim 0.97$. The observed acoustic
peaks~\cite{Bennett+etal:2003a} at later times will follow from our early universe
prediction of the spectrum by an analysis similar to the one used in inflation
theory~\cite{Liddle+Lyth:2000}. We have yet to derive the tensor modes spectrum in
our model, and a derivation of the polarization of the spectral modes is also needed
to compare with the recent WMAP data~\cite{Bennett+etal:2003b}. Since we have truly
primordial matter in this model, we must also be cautious about the possibility of
radiation structure re-emerging and dominating the scalar modes at later times. One
might expect that the enlarged VSL light cone can produce Cherenkov radiation that
could dilute this source of radiation. This is an issue that requires further
consideration.

We have set out to show that a purely VSL motivated scheme, within the
scenario of the scalar-tensor bimetric gravity theory, can successfully
lead to a scale invariant CMB spectrum. By using the Hollands-Wald idea of
having the perturbative scalar wave modes born in a spatial volume {\it
outside} the horizon whose size is governed by the length scale
$\sqrt{12B}$, we have demonstrated that we have indeed
succeeded in our quest. The origin of the spacetime foam region which
determines the ground state of the perturbative modes still needs to be
understood, but the model can be considered a self-consistent way to explain
the origins of the seeds of galaxy formation which is not based on a standard
inflationary model.

\section*{acknowledgments}

One of the authors (JWM) thanks Glen Starkman for enlightening discussions; this research was supported by the Natural Science and Engineering Research Council of Canada.


\end{document}